\documentclass[11pt,twoside]{article}

\usepackage{paperlighter}

\usepackage{microtype}
\usepackage{graphicx}
\usepackage{subfigure}
\usepackage{booktabs}

\usepackage{amsmath}
\usepackage{amssymb}
\usepackage{mathtools}
\usepackage{amsthm}

\usepackage{apacite}

\usepackage[capitalize,noabbrev]{cleveref}

\usepackage{pgfplots}
\pgfplotsset{compat=1.18}
\usepackage{tikz}

\slimtitle{Designing Human-AI Systems: Anthropomorphism and Framing Bias on Human-AI Collaboration}
\slimauthor{Samuel Aleksander Sánchez Olszewski}

\begin{document}
\hspace{-32pt}
\lightertitle{Designing Human-AI Systems: Anthropomorphism and Framing Bias on Human-AI Collaboration}

\lighterauthor{\hspace{-24pt} Samuel Aleksander Sánchez Olszewski$^{*}$}

\lighteraddress{$^*$}{Universidad Carlos III de Madrid, 28903 Getafe, Spain}

\lighteremail{samuelaleksander.sanchez@alumnos.uc3m.es}

\begin{abstract}
AI is redefining how humans interact with technology, leading to a synergetic collaboration between the two. Nevertheless, the effects of human cognition on this collaboration remain unclear. This study investigates the implications of two cognitive biases, anthropomorphism and framing effect, on human-AI collaboration within a hiring setting. Subjects were asked to select job candidates with the help of an AI-powered recommendation tool. The tool was manipulated in a 3 x 3 between-subjects design to present three different AI identities (human-like, robot-like, generic) and three types of framing (positive, negative, and neutral). The results revealed that the framing of AI’s recommendations had no significant influence on subjects’ decisions. In contrast, anthropomorphism significantly affected subjects’ agreement with AI recommendations. Subjects were less likely to agree with the AI if it had human-like characteristics. These findings demonstrate that cognitive biases can impact human-AI collaboration and highlight the need for tailored approaches to AI product design, rather than a single, universal solution.

\end{abstract}

\section{Introduction}

\begin{figure}[ht]
\vskip 0.2in
\begin{center}
\centerline{\includegraphics[width=0.7\columnwidth]{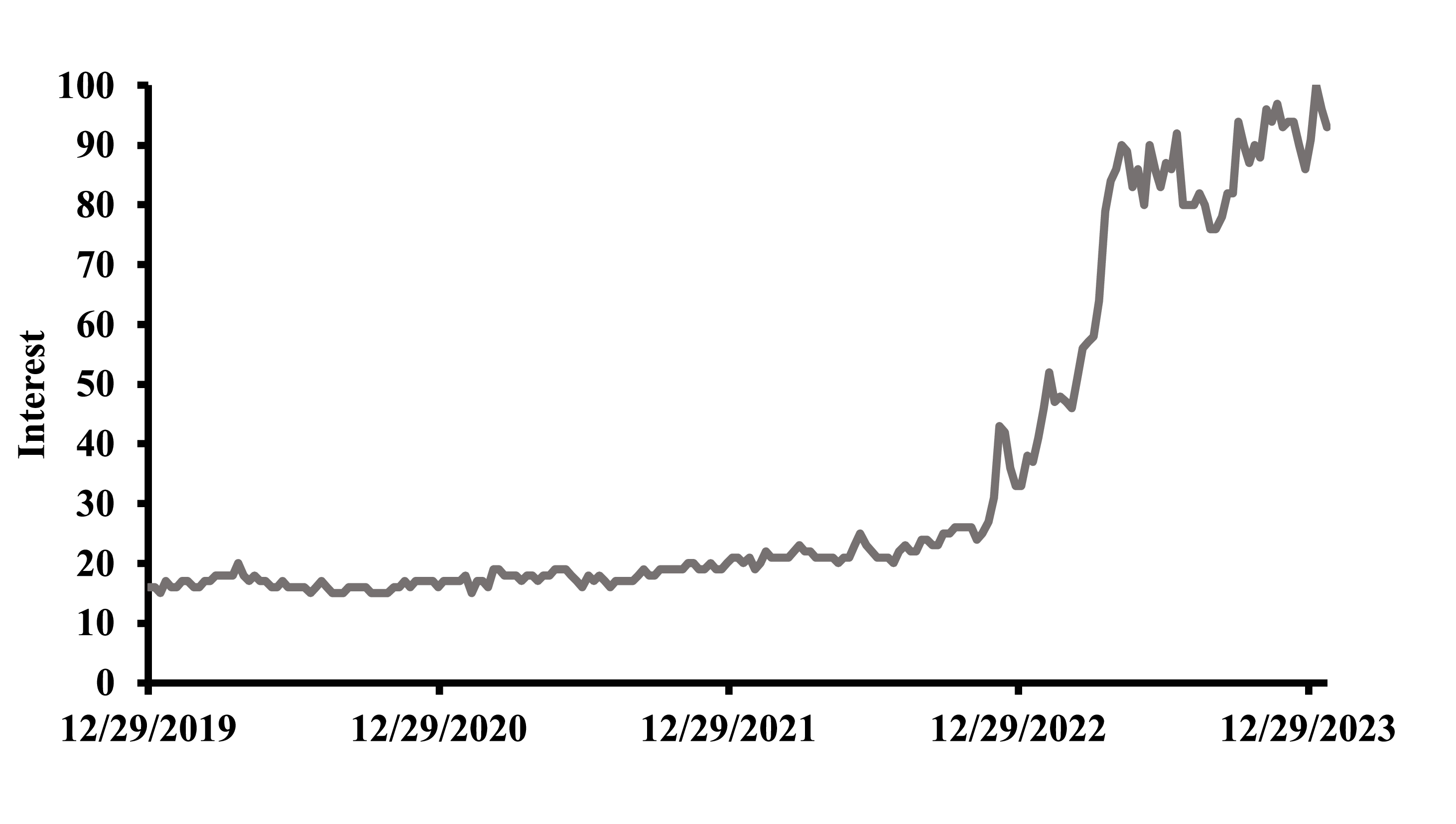}}
\caption{Own elaboration based on Google Trends data for the interest in the term ‘AI’ from 12/29/2019 to 01/27/24. Numbers represent search interest relative to the highest point. A value close to 100 indicates that we are currently at the peak popularity of the term. }
\label{AI term}
\end{center}
\vskip -0.2in
\end{figure}

The rapid rate at which Artificial Intelligence (AI) technologies have been embraced worldwide during the last months suggests that we are in the midst of an AI boom. Google Search Trends data indicates that since the release of OpenAI’s chatbot, ChatGPT, on the 30th of November of 2022 \cite{openai_introducing_2022}, interest in AI has grown significantly (see \cref{AI term}). Likewise, according to a recent study, ChatGPT reached 123 million monthly unique visitors in less than three months, making it the fastest-growing internet application of the last 20 years \cite{r_chow_how_2023}. 

The current momentum of AI adoption has raised concerns about the future of work. While some argue that AI has the disruptive potential to replace humans in the workplace, the complementary characteristics of humans and AI suggest that we are moving towards a synergetic collaboration between the two, where AI augments human capabilities and vice versa \cite{jarrahi_artificial_2018}. However, research on the role of human perception in this collaboration remains overlooked \cite{rastogi_deciding_2022}.

This study investigates the impact of two cognitive biases, anthropomorphism and framing effect, on human-AI collaboration in the context of hiring decision-making. An experiment was designed to simulate the screening phase of the recruitment process, where companies are already using AI-based tools (see \cref{sec: hiring process} and \cref{sec: AI-enabled recruiting}). This study aims to expand our understanding of the cognitive consequences of human-AI collaboration and contribute to the design of AI-based products that consider these cognitive factors.

\section{Related work}
The synergy between AI and human capabilities seems to be leading to a transition from individual interaction to collaborative work, which could enhance AI’s advantages. Some studies have already started to explore the potential of human-AI collaboration, with results suggesting that it can increase human productivity. For instance, a study by \citet{sowa_digital_2020} found that participants who collaborated with a virtual assistant during a business problem simulation achieved higher productivity and reported being more satisfied with their performance than those who did the tasks on their own. 

The role of technology is shifting from a mere aid for completing tasks to a collaborative partner \citep{grimm_team_2018, sowa_cobots_2021}. AI’s unique capabilities lead to new challenges and opportunities, fomenting a reevaluation of Human-Computer Interaction frameworks and methodologies when it comes to AI. Consequently, researchers propose transitioning from the human-centered design philosophy that has guided Human-Computer Interaction in favor of a Human-Centered AI strategy, paving the way for the emerging field of Human-AI Interaction \citep{wang_human_human_2020}.

A recurring theme among the different Human-AI Interaction frameworks is prioritizing humans during AI system development (see for example: \citealp[]{shneiderman_bridging_2020}; \citealp[]{xu_toward_2019}). So far, AI research has primarily focused on solving technical problems rather than addressing human needs, resulting in a “technology-centered design” of AI systems \citep{wang_human_human_2020,xu_toward_2019}. One way to center AI development on humans and enhance collaboration is by creating shared mental models that facilitate a deeper comprehension of each other’s capabilities. For AI and humans to understand each other, humans must be able to interpret AI outputs accurately, and AI agents should be designed to anticipate human interpretation \citep{kaur_building_2019}. 

To ensure that AI can comprehend human interpretation, it is essential to integrate considerations of human cognition into the development of AI systems. Human biases are a crucial component of human cognition, yet research on their implications in human-AI collaboration is scarce \citep{rastogi_deciding_2022}. The present study explores how we should design AI tools by investigating cognitive biases and their consequences in human-AI collaboration.

\subsection{Cognitive bias and dual process theories}

While the study of heuristics dates back several centuries, the work of Daniel Kahneman and Amos Tversky, inspired by Herbert Simon's ideas, marked the beginning of a new era in the study of biases in human judgment and decision-making \citep{hjeij_brief_2023}. However, the concept of heuristics is still controversial \citep{scott_heuristics_2015}. In this work, we adopt Kahneman and Tversky’s definition of heuristics as mental methods that help humans make quick judgments under uncertainty. These heuristics often lead to errors in decision-making, which are referred to as cognitive biases \citep{tversky_judgment_1974}.    

Tversky and Kahneman launched the heuristics and biases program in the early 1970s \citep{tversky_belief_1971}, and shared their main findings in the influential article “Judgement Under Uncertainty: Heuristics and Biases” (\citeyear{tversky_judgment_1974}). The article documented a collection of cognitive biases caused by the reliance on heuristics and provided empirical evidence for their existence. The influence of this article has been substantial, remaining more than 40 years later as one of the most frequently cited works in social science literature \citep{kahneman_thinking_2011-1}.  
The heuristics and biases program of Tversky and Kahneman falls within a long-standing tradition that characterizes human cognition as involving two complementary forms of processing. According to these dual-process theories, we can differentiate two types of cognitive operations. One of them, referred to as type 1, is fast, automatic, and non-conscious. The other one, type 2, is slow, controlled, and conscious \citep{evans_dual-process_2013}. Type 1 processes enable us to perform tasks effortlessly, whereas type 2 is responsible for complex and deliberate thinking \citep{kahneman_thinking_2011-1}. The body of literature that builds upon the heuristics and biases program associates heuristics and the resulting cognitive biases with the operation of the automatic type 1 processes \citep{gilovich_representativeness_2002}. 

\subsection{The hiring process}
\label{sec: hiring process}

The impact of Kahneman and Tversky’s work extends far beyond the field of psychology \citep{kahneman_thinking_2011-1}. During the last 40 years, the literature that analyzes heuristics and cognitive biases of business managers has experienced a significant surge, with a particularly notable increase over the last decade. Due to the role that business leaders play in shaping companies’ strategy, understanding how cognitive biases may influence their decision-making process is crucial for explaining organizational outcomes \citep{hodgkinson_heuristics_2023}.

One of the most critical decisions that business leaders need to make is deciding who to hire \citep{schmitt_selection_2007}. Nevertheless, hiring requires considerable time and resources. An inefficient hiring strategy may lead to prolonged delays that could deter potential candidates \citep{kim-schmid_where_2022}. In a world where talent acquisition is becoming a strategic business priority, AI-enabled recruiting has established itself as a critical objective for companies \citep{van_esch_factors_2019}. In addition, the need to mitigate human bias (e.g., racial, ethnic, or gender bias) is accelerating developers’ efforts to design unbiased AI systems that may assist during the hiring process \citep{yarger_algorithmic_2019}. 

The hiring process can be conceptualized as a “compound decision process”, or “pipeline”, that is comprised of multiple stages \citep{bower_fair_2017}. Companies may have slightly different variations of these stages, but they typically comprise the following four: sourcing, screening, interviewing, and selection. The sourcing stage involves attracting potential candidates through various means, such as advertisements or job postings. During the screening stage, recruiters evaluate candidates’ qualifications and determine which candidates should proceed to the interview stage. Next, in the interviewing stage, recruiters conduct a more personalized assessment of candidates. Finally, during the selection stage, recruiters reach a final hiring decision \citep{miranda_bogen_help_2018}.

This study focuses on the screening stage of the hiring process. In this stage, companies are already using AI tools to assist their decision-making process regarding candidates’ assessments. Some of these tools review applicants’ resumes and rate them based on a comparison between the resumes and the recruiter’s required qualifications \citep{miranda_bogen_help_2018}. 

\subsection{AI-enabled recruiting and cognitive biases}
\label{sec: AI-enabled recruiting}

As it has been mentioned previously, companies are already taking advantage of AI-enabled recruiting to make the hiring process more efficient and ultimately increase their firm’s value \citep{van_esch_factors_2019}. However, the adoption of AI applications in recruitment is also raising concerns about its ethical implications \citep{hunkenschroer_ethics_2022}. For instance, in 2018, a resume screening service developed and used internally by Amazon showed gender bias \citep{dastin_amazon_2018}.

Even though companies are already embracing AI, research about the presence of bias in both human and AI decision-making processes during recruitment remains limited \citep{silva_human_2023}. Notably, one area that has received scant attention is the cognitive bias that may arise in the interaction between humans and AI. While a few studies have investigated the impact of AI-powered recruiting tools on decision-making, their focus has primarily been on the phenomenon of algorithmic bias \citep{hunkenschroer_ethics_2022}. In fact, in their literature review on the ethical implications of AI-enabled recruiting, \citet{hunkenschroer_ethics_2022} did not identify any studies that addressed cognitive biases. Likewise, \citet{hodgkinson_heuristics_2023} analysis of literature on heuristics and biases of top managers did not mention any relevant article related to strategic decision-making during the recruitment process. 

The notion that AI will eventually replace humans in the workplace has become a widely held belief. While it is true that automation and algorithmic systems are becoming more prevalent in Human Resources (HR) departments, human decision-making remains a crucial component in high-stakes choices. AI tools are commonly used to aid recruiters in rating or ranking job applicants, but it is ultimately up to human recruiters to make the final decisions regarding candidate selection \citep{fernandez-martinez_ai_2020}. Consequently, it is fundamental to understand how cognitive bias may affect the interaction between the human and the AI. 

\subsection{Cognitive biases and Human-AI Collaboration}

This work examines how an AI decision-aid tool might influence the recruiter’s decision-making process when AI’s recommendations show two types of biases: anthropomorphism and the framing effect. This study should provide us with more information about the effect of cognitive biases on human-AI interaction. Ultimately, it tests the following two hypotheses:

\begin{itemize}
    \item[-] H1: Framing of the AI recommendation will influence the degree to which the human conforms to it (framing effect). 
    \item[-] H2: Humans are more likely to follow AI instructions if it has a human-like identity (anthropomorphism). 
\end{itemize}

\subsubsection{Anthropomorphism and trust}

By anthropomorphism, we refer to the tendency to ascribe human traits to non-human entities, in the context of social interactions, to rationalize their behavior \citep{duffy_anthropomorphism_2003}. Anthropomorphism can be considered a cognitive bias since it consists of an erroneous human perception caused by an unconscious process \citep{dacey_anthropomorphism_2017}. That is to say, it is associated with the heuristic and automatic type 1 processes. 

Researchers have already explored the potential of intentionally incorporating human-like features (anthropomorphic design) to amplify existing tendencies toward anthropomorphism. This approach aims to enhance human interaction with technology, particularly in fields like human-robot interaction \citep{fink_anthropomorphism_2012}. Anthropomorphic designs could also facilitate human-AI interaction in various ways \citep{duffy_anthropomorphism_2003}, one of which is by increasing human trust in AI. According to previous research, the extent to which humans are willing to depend on automated systems in uncertain scenarios is largely influenced by trust \citep{hoff_trust_2015}. Moreover, trust plays a critical role in mitigating the uncertainty associated with the use of AI among humans, thereby determining the acceptance and adoption of such technologies \citep{lukyanenko_trust_2022}.

In a seminal study, researchers observed a higher level of trust when assigning human-like attributes to autonomous vehicles. Subjects who drove a vehicle that had a name, a gender, and a voice reported more trust in their vehicles than those who drove the same car but without humanlike features \citep{waytz_mind_2014-1}. Likewise, another study demonstrated that subjects who used a smartphone app with the image of a doctor to assist them in diabetes management decisions trusted the system more than those who used the same app with the image of gears \citep{pak_decision_2012}.

In addition, anthropomorphism can have important implications for product development. It has been observed as a factor that increases the adoption of personal intelligent agents \citep{moussawi_how_2021}. Additionally, \citet{sidlauskiene_correction_2023-1} found that subjects perceived products as more personalized and were willing to pay a higher price for them if they interacted with a chatbot that had human-like features. 

Regardless of the aforementioned cases, few studies have investigated how anthropomorphism can affect human trust in automatic systems that assist in decision-making \citep{pak_decision_2012}. Additionally, the literature on how the anthropomorphic design of AI influences user trust remains scarce \citep{troshani_we_2021}. Although companies are already using AI tools to assist their hiring processes, no research has been found on how their anthropomorphic design could affect recruiters' trust. 

A non-significant association between anthropomorphic design and human trust may suggest that subjects are already anthropomorphizing the AI during its interactions, as some previous studies suggest \citep{chinmulgund_anthropomorphism_2023}. Likewise, a negative association could imply that the attribution of human-like characteristics to an AI agent is actually decreasing human affinity, which might be related to the uncanny valley effect (see \citealp[]{mori_uncanny_1970}). On the other hand, a positive association might indicate that providing human traits to an AI makes subjects view the AI as more human.

\subsubsection{Framing effect}

The concept of “framing effect” refers to the tendency of decision-makers to respond differently when presented with distinct (yet equivalent) descriptions of the same problem \citep{frisch_reasons_1993}. Although framing effects tend to be treated homogeneously, we can distinguish three different types of framing effects: risky choice framing, goal framing, and attribute framing \citep{levin_all_1998}. Risky choice framing has often been regarded as the prototypical framing effect due to the repercussion of the “Asian disease problem”, which was introduced by \citet{tversky_framing_1981}. In this experiment, subjects shifted from risk aversion to risk-taking attitudes depending on the framing of the outcome. When the outcome of a decision was described by the number of lives saved (positive frame), the majority choice was risk averse. On the contrary, if the outcome was described by the number of lives lost (negative frame), the majority choice was risk-taking. Goal framing measures how subjects’ decisions deviate depending on how the consequences of their actions are framed. In the positively framed condition, only the potential benefits of achieving a specific result are shown to participants whereas in the negatively framed conditions, the focus is shifted to the negative consequences that would be avoided by that result \citep{levin_all_1998}.

The present study focuses on attribute framing, which has also been extensively studied in the literature. Attribute framing addresses how the description of a single attribute of an object can impact decision-maker preferences \citep{levin_all_1998}. The attribute is typically described in two different ways, using a positive or a negative frame. The evidence indicates that individuals evaluate the object more favorably when the positive frame is presented to them \citep{leong_role_2017}. For example, a study found out that job candidates who are exposed during realistic job previews to positively framed messages (e.g., “Eighty-five percent of the employees who do this job are satisfied with the job”) have higher job expectations and willingness to accept a job offer than those that were given the same information with a negative frame (e.g., “Fifteen percent of the employees who do this job are dissatisfied with the job”), \citep{buda_message_2003}.

In a recent study, \citet{jin_how_2017-1} examined how online consumers modify their purchasing decisions depending on the framing of product attribute descriptions. Their results showed that consumers under the positive framing condition had a higher purchasing intention compared to the negative framing condition. Moreover, negative messages led to increased cognitive conflict and difficulty in decision-making. These findings could have practical implications for the development of AI-based products. For instance, AI chatbots on websites might be instructed by companies to present the product’s attributes using positive frames.

Nevertheless, the use of AI tools also raises ethical concerns, particularly in certain domains, such as the recruitment sector. Inconsistent use of positive and negative frames in AI screening tools may lead to candidate discrimination. AI-enabled recruiting tools are already adopted by organizations, and due to their current design, recruiters may be subject to framing effects. AI-powered screening software\footnote{See for example Ideal, a tech company acquired by Ceridian which developed an AI tool that allows recruiters to “instantly screen and shortlist thousands of candidates.”} typically grades and ranks candidates based on their predicted fit for the role. These tools provide recruiters with the relative position of a specific candidate within the pool of applicants (e.g., “There is a 90\% probability of hiring someone worse than this candidate”). Note that the same information could be framed differently. It is equivalent to saying that “there is a 10\% probability of hiring someone as good or better than this candidate.” 

Previous research on the framing effect suggests that the frame used by an AI tool could potentially affect the recruiter’s perception of the candidate. If this is the case, it would be desirable to design AI tools where the framing effect is taken into account in order to avoid candidates’ discrimination. To the extent that humans are susceptible to cognitive biases, the decision-aid technologies we develop should be designed in a way that mitigates these biases 
\citep{arkes_costs_1991}. For example, ensuring that AI tools display two different frames or the same frame for all the different candidates. 

Likewise, AI-enabled recruiting tools could play an important role in reducing the framing effect on recruiters. Since resumes are created by candidates in a non-standardized fashion, resumes of different candidates may display frames of opposed valence. We could therefore think of a hypothetical scenario where two candidates with the same credentials present themselves using different frames. The framing chosen by the candidates may be a factor that influences the recruiter’s final decision. To mitigate the framing effect, AI tools could extract data from candidates’ resumes and standardize it so that information is displayed using a frame with the same valence. 

\section{Experiment}

For this study, we designed a binary prediction task within a recruitment setting to address two main research questions. First, we build on the work of \citep{rastogi_deciding_2022} and investigate the impact of cognitive bias in the context of AI-assisted decision-making. Particularly, we focus on attribute framing by measuring the effect of positively and negatively framed AI recommendations. Second, we aim to understand how AI appearance (either human-like or robot-like) affects human trust, following previous work on the subject \citep{pak_decision_2012}. 

\subsection{Participants}

To run the experiment, 750 participants from Prolific were recruited. Prolific is an online platform that allows researchers to recruit participants from a diverse pool. Since the survey was in English, the pool was limited to English native speakers from the United States and the United Kingdom. Participants were compensated £7.5 per hour, which the platform regarded as fair based on several factors, such as the complexity of the task or the amount of effort required.

\subsection{Task and AI model}

The task aimed to emulate the screening phase of a recruitment process within a private company. Participants were asked to act as recruiters and decide whether a student should be invited for an interview or not. To help them in their decisions, we provided some details about the candidates (e.g., grades, work experience, area of specialization), and an AI recommendation.  

The dataset used provided information about 215 candidates, including the aforementioned details and their employment situation (“placed” or “not placed”). The dataset and the methodology employed to generate the AI recommendations were directly sourced from the discussions and resources available on the Campus Recruitment contest on Kaggle.com \citep{noauthor_campus_2020}. After handling missing values, we preprocessed the data by encoding the categorical variables into numerical values and splitting the data into training and testing sets (80:20 split). Then, we fit a logistic regression model on the dataset, obtaining an accuracy rate of 83\%. Additionally, we added a new column with the predicted probability for each candidate within the test set. These predicted probabilities were used for the AI recommendation and were unique per candidate. However, the form in which they were presented to subjects varied, as will be discussed in another section. 

To create a more realistic scenario, we briefly described the company and the position in the instructions part before the actual survey. Furthermore, a name was attributed to each participant. To avoid possible discriminations that could invalidate the results, we tried to select names as neutral as possible. An equal distribution between names of diverse gender identities was also pursued. 

\subsection{Study Procedure}

\begin{figure}[ht]
\vskip 0.2in
\begin{center}
\centerline{\includegraphics[width=0.5\columnwidth]{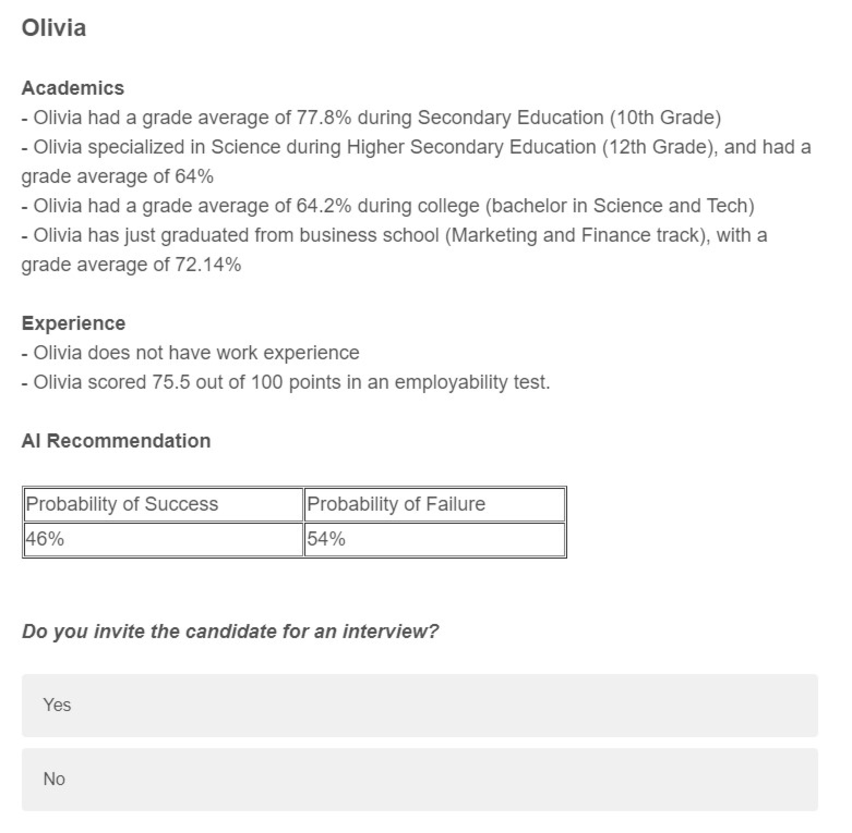}}
\caption{Slide of a candidate from the training section with its three main sections: Academics, Experience, and AI Recommendation. Example of a subject exposed to the neutral condition of both treatments.}
\label{training-section}
\end{center}
\vskip -0.2in
\end{figure}

The survey was made on Qualtrics, and it had fifteen questions. Every question presented a distinct candidate with different characteristics. In each of these questions, subjects were asked to decide whether to invite the candidate to the interview or not. The questions had identical structure and predictive tasks. They included a section on the academic details, another one on the employability details, and a final one with the AI recommendation (see \cref{training-section}). AI's recommendation section was the only one that changed among subjects.

To emulate the recruiter’s prior experience and enhance the realism of the survey, we included a training section consisting of five questions. Candidates in this section were identical for all subjects, but they were displayed in a random order. The candidates were selected so that they covered all the ranges from [0.5, 0.6] to (0.9, 1] of the model’s estimated probability. Using these estimations as a proxy for difficulty, this selection ensured that the training section contained diverse levels of difficulty \citep{rastogi_deciding_2022}.

\begin{figure}[ht]
\vskip 0.2in
\begin{center}
\centerline{\includegraphics[width=0.5\columnwidth]{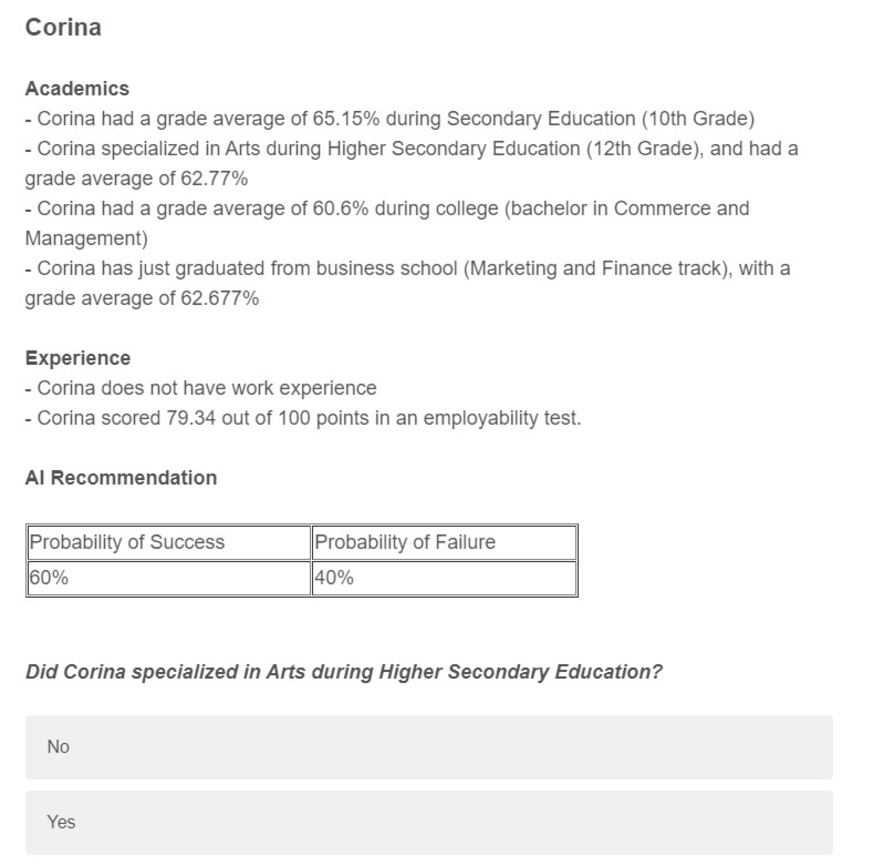}}
\caption{Example of one of the two attention checks. Although the structure of the slide remains the same, the question changes. Instead of the question: “Do you invite the candidate for an interview”, subjects are now asked “Did Corina specialize in Arts during Higher Secondary Education?"}
\label{attention check}
\end{center}
\vskip -0.2in
\end{figure}

Once subjects completed the training section, they moved to the test set, consisting of ten questions. As in the previous section, the candidates presented to the subjects were the same for all, but the order was randomized. In addition to these questions, two different attention checks were included at the end of each section (see \cref{attention check}). 

\subsection{Experimental design and hypotheses}

The experiment followed a factorial design of two variables and three treatment levels \citep{millsap_experimental_2009}. It had a 3 x 3 between-subjects design, with three different AI identities (human-like, robot-like, generic) and three types of framing (positive, negative, and neutral). Once a subject started the survey, a specific number from one to nine was randomly assigned to them. Each number had a corresponding treatment combination assigned and remained the same for participants during the whole survey flow. The dependent variable for the two hypotheses was the correlation between the AI recommendation and the subject’s decision, indicating the frequency of agreement or disagreement between the two.

\subsubsection{Hypothesis 1 (H1)}

To validate the first hypothesis (H1), three different variations of the AI recommendation were designed. These variations distinguished between each other in how the predicted probability estimated by the model was framed (positive, negative, or neutral frame). In the positive frame, it was described in terms of success rates (e.g., “80\% probability of success”), while in the negative frame, the same probability was described in terms of failure rates (e.g., “20\% probability of failure”). 

\begin{figure}[ht]
\vskip 0.2in
\begin{center}
\centerline{\includegraphics[width=0.5\columnwidth]{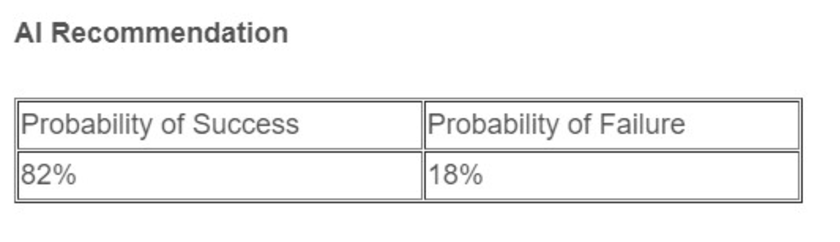}}
\caption{An example of a neutral frame with positively framed information on the first column.}
\label{neutral 1}
\end{center}
\vskip -0.2in
\end{figure}

\begin{figure}[ht]
\vskip 0.2in
\begin{center}
\centerline{\includegraphics[width=0.5\columnwidth]{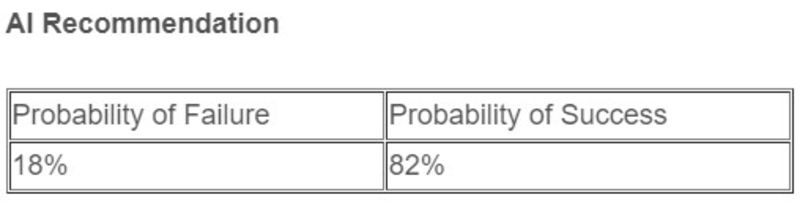}}
\caption{An example of a neutral frame with negatively framed information on the first column.}
\label{neutral 2}
\end{center}
\vskip -0.2in
\end{figure}

Note that equivalent frames could have been achieved with different descriptors: we could have labeled the positive outcome with a negative descriptor (“80\% probability of not being a failure”), and the negative outcome with a positive descriptor (“20\% probability of not being a success”). However, unlike outcome-framing manipulation, descriptors do not seem to have a significant impact on attribute-framing bias. Therefore, we opted for a balanced presentation of both positive and negative framings in the neutral treatment level to minimize the framing effect, as prior work on the subject suggests \citep{kreiner_alive_2019}. To further ensure neutrality in the neutral treatment level, descriptors, and the estimated probabilities were displayed in a table. Moreover, we designed two variants of the table with distinct frame order: one table had the positively framed information in the first column (\cref{neutral 1}), and the other table placed the same information in the second column (\cref{neutral 2}). To neutralize unintended effects due to the ordering of the columns, subjects exposed to the neutral framing were randomly assigned to one of these two variations.

To verify frames’ valence, we corroborated the polarity of the words “success” and “failure” with the Subjectivity Lexicon presented in the work “Recognizing Contextual Polarity in Phrase-Level Sentiment Analysis” \citep{wilson_recognizing_2005-1}. According to the Lexicon, these words display a positive and negative polarity out of context, and these polarities are considered weakly subjective. 

\subsubsection{Hypothesis 2 (H2)}

In the second hypothesis (H2) we wanted to measure the impact of AI’s anthropomorphic design on human trust. To test the hypothesis, we designed three different AI identities, which differed in the nature of the features attributed to the AI (human-like, robot-like, and generic). In the human-like condition, the AI was given humanlike features: it was referred to by a human name (Maya), was given a gender (female), and was given a human appearance (using an image of a real human). In the robot-like condition, the AI had a robotic name (ZAX-D2), an impersonal pronoun (it), and was represented with a humanoid robot image. For the generic condition (the control group), we omitted any reference to its identity by replacing the names given in the robot-like and human-like conditions with the word “AI”. In addition, no picture was provided. 

\begin{figure}[ht]
\vskip 0.2in
\begin{center}
\centerline{\includegraphics[width=0.5\columnwidth]{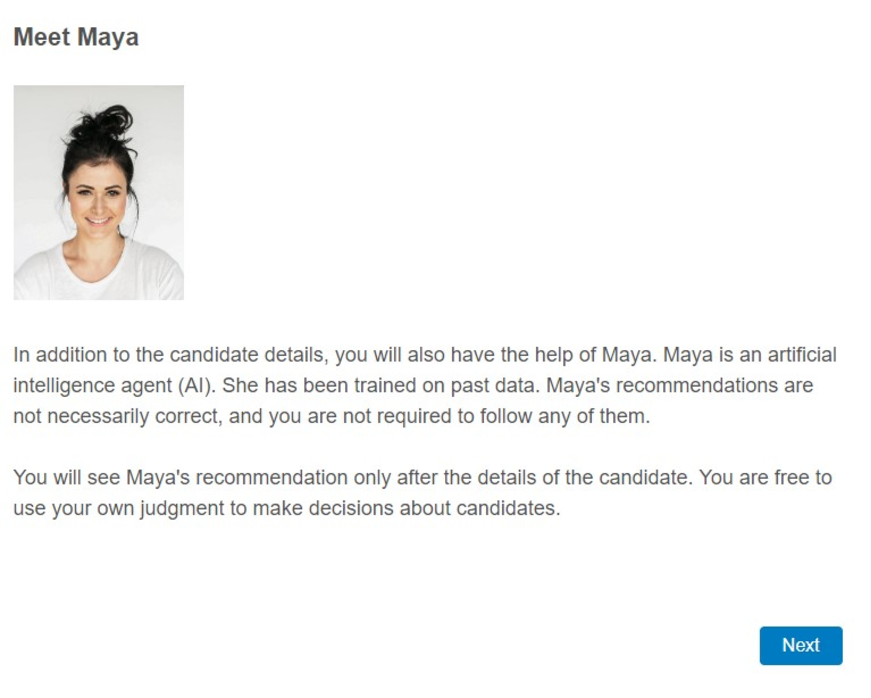}}
\caption{Introduction provided to subjects exposed to the human-like condition (Maya).}
\label{Intro-human}
\end{center}
\vskip -0.2in
\end{figure}

\begin{figure}[ht]
\vskip 0.2in
\begin{center}
\centerline{\includegraphics[width=0.5\columnwidth]{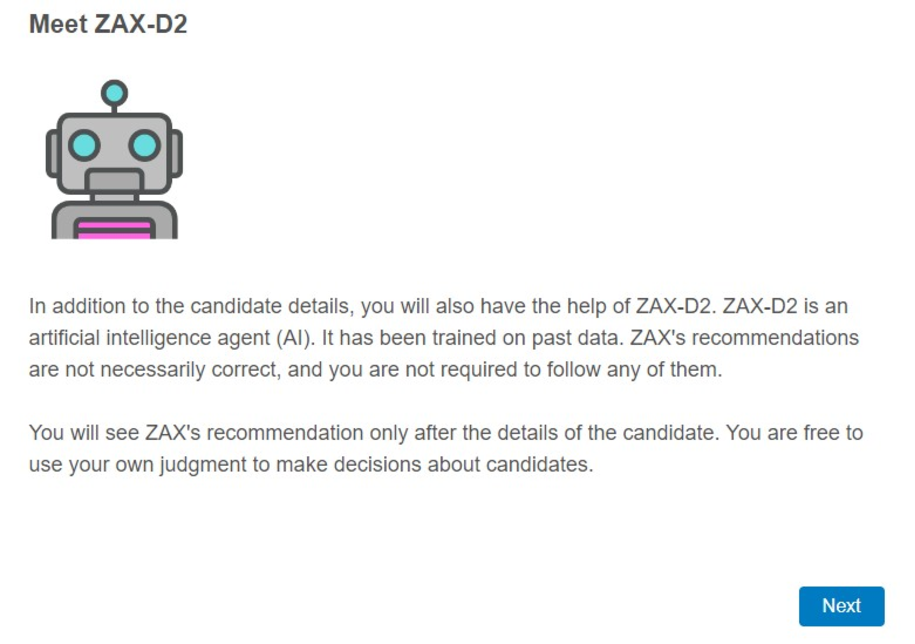}}
\caption{Introduction provided to subjects exposed to the robot-like condition (ZAX-D2).}
\label{Intro-robot}
\end{center}
\vskip -0.2in
\end{figure}

Subjects were exposed to AI’s anthropomorphic design in two different parts of the survey. Their first contact occurred in the instructions part. A specific slide was made to introduce AI to subjects (see \cref{Intro-human,Intro-robot}). Then, AI’s anthropomorphic features were also present in the AI recommendation section of every question: the name given appeared in the header of the section (e.g., “Maya recommendation”, “ZAX-D2 recommendation”, or “AI recommendation”), along with the correspondent picture (if applicable). 

It should be mentioned that there are many ways in which human-like and robot-like features can be attributed to AI. We are aware of the fact that more complex cues could be introduced to enhance the anthropomorphic effect. For instance, \citet{schanke_estimating_2021} used humor, social presence, and communication delays to induce anthropomorphism in the interaction with chatbots. However, these cues might require the implementation of a more sophisticated interface. 

The specific number and nature of human-like characteristics required to elicit perceptions of a system as humanlike remain unclear \citep{de_visser_almost_2016}. In response to this uncertainty, we adopted an anthropomorphic design similar to the one proposed by \citet{pak_decision_2012}, which reported the influence of human-like clues in subjects' trust through the use of images. Additionally, we integrated subtle linguistic cues, such as the use of pronouns (e.g., “she” vs “it”). 

Example slides of both framing (positive and negative), and anthropomorphic (human-like and robot-like) conditions are available in the \cref{appendix}.

\subsection{Results}

To investigate differences in subjects’ agreement rates with AI recommendations as a function of framing and AI identity, a 3 (framing: positive, negative, neutral) X 3 (AI identity: human-like, robot-like, generic) ANOVA with unbalanced replications was run. Since a large interaction between the two factors was not expected, we used a Type II ANOVA \citep{smith_factorial_2021}. We assumed subjects agreed with the AI if they invited candidates to which the AI assigned a probability of success greater than or equal to 50\% or if they declined to invite a candidate to which the AI assigned a probability of success smaller than 50\%. To calculate agreement rates, we counted the number of times a subject agreed with the AI and divided it by the total number of questions.

As shown in \cref{attention check performance}, a considerable percentage of subjects failed at least one attention check. Dropping subjects who failed these attention checks could potentially cause some unintended biases \citep{aronow_note_2019}. To mitigate this effect, we decided to run the ANOVA three times using different sample sizes based on attention check performance. The first ANOVA iteration included all subjects exposed to a specific treatment condition, regardless of the number of attention checks passed (\(a \geq 0\)). The second iteration considered subjects that passed at least one attention check (\(a \geq 1\)). Finally, the third iteration included only subjects who passed the two attention checks (\(a = 2\)).

\begin{table}[H]
\centering
\begin{tabular}{|l|c|c|c|c|c|c|}
\hline
 & \multicolumn{3}{c|}{\textbf{AI Identity}} & \multicolumn{3}{c|}{\textbf{Framing}} \\
\hline
 & \textbf{Human} & \textbf{Robot} & \textbf{Generic} & \textbf{Positive} & \textbf{Negative} & \textbf{Neutral} \\
\hline
Total nº of subjects (\(a \geq 0\)) & 247 & 249 & 253 & 243 & 258 & 248 \\
Passed an attention check (\(a \geq 1\)) & 223 & 230 & 222 & 206 & 239 & 230 \\
Passed 2 attention checks (\(a = 2\)) & 126 & 121 & 113 & 103 & 143 & 114 \\
\hline
\end{tabular}
\caption{Subjects’ attention check performance by treatment group.}
\label{attention check performance}
\end{table}

\cref{summary of ANOVA results} presents the ANOVA results for the three different sample sizes (\(a \geq 0\), \(a \geq 1\), \(a = 2\)). It includes the F-statistics, p-values, and their corresponding significance levels for each factor and their interactions. These results did not support the hypothesized relationship between framing and agreement rates; no statistically significant differences in agreement rates were found between subjects exposed to different framings, regardless of their attention check performance. Interaction effects did not have a statistically significant impact either. However, there was a significant effect of AI identities on agreement rates, \( F(2, 357) = 2.554, \; p\text{-value} = 0.079 \), for subjects who passed both attention checks.

\begin{table}[H]
  \centering
  \setlength{\tabcolsep}{4pt} 
  \begin{tabular}{|l|c|c|c|c|c|c|c|c|c|}
    \hline
    & \multicolumn{3}{c|}{\textbf{Framing}} & \multicolumn{3}{c|}{\textbf{AI Identity}} & \multicolumn{3}{c|}{\textbf{Interaction}}  \\
    \hline
    & \(a \geq 0\) & \(a \geq 1\) & \(a = 2\) & \(a \geq 0\) & \(a \geq 1\) & \(a = 2\)  & \(a \geq 0\) & \(a \geq 1\) & \(a = 2\) \\
    F-statistic& 1.183& 0.481& 0.504& 1.488& 1.448& 2.554& 1.186& 1.176& 1.547\\
    p-value & 0.307& 0.618& 0.604& 0.226& 0.236& 0.079& 0.316& 0.32& 0.188\\
    significance & - & - & - & - & - & +& - & -& -\\
    \hline
  \end{tabular}
  \caption{Summary of the Type II ANOVA results for the two different factors (AI identity, Framing) and their interactions. The ANOVA was run three times with different sample sizes, categorized by attention check performance. The table includes the F-statistics, the p-values, and their corresponding significance levels. Treatments with a p-value greater than 0.1 are indicated by a hyphen ("-"), whereas treatments with a p-value less than 0.1 are indicated by a plus sign ("+").}
  \label{summary of ANOVA results}
\end{table}

To further investigate the effect of AI identities on subjects who passed both attention checks, we conducted a post-hoc analysis using the Tukey-Kramer method to account for the unbalanced design of the experiment. As shown in \cref{HSD: mean_differences}, there is a significant difference in agreement rates between the generic and human identity conditions. Subjects exposed to the human condition were on average 3.99\% less likely to agree with the AI in any given question compared to the control group, with a significance level of \(p = 0.0987 \). The study's findings indicate that attributing a human-like identity to the AI significantly impacts agreement rates. However, contrary to our expectations, the results show that subjects are more likely to agree with the AI's recommendations when it lacks a human identity.

\begin{figure}
\centering
\begin{tikzpicture}
\begin{axis}[
    width=12cm,
    height=8cm,
    xlabel={Mean Difference},
    ytick={1,2,3},
    yticklabels={Human - Robot, Generic - Robot, Generic - Human},
    xmin=-0.1, xmax=0.1,
    xtick={-0.08, -0.04, 0, 0.04, 0.08},
    xticklabel style={/pgf/number format/fixed, /pgf/number format/precision=2},
    y dir=reverse,
    ylabel near ticks,
    xlabel near ticks,
    enlarge x limits=0.1,  
]

\draw[dashed] (axis cs:0,0) -- (axis cs:0,3.5);

\addplot[only marks, error bars/.cd, x dir=both, x explicit]
    coordinates {
    (0.0037, 1) +- (0.0391,0)
    (-0.0362, 2) +- (0.0401,0)
    (-0.0399, 3) +- (0.0398,0)
};

\end{axis}
\end{tikzpicture}
\caption{Mean differences between AI identities with 90\% confidence intervals based on Tukey HSD test. The dashed vertical line represents zero difference. Intervals that do not cross this line indicate significant differences (\(p < 0.10\)).}
\label{HSD: mean_differences}
\end{figure}

In sum, none of the stated hypotheses has been validated by the experiment. The framing of the AI recommendations did not significantly impact the agreement rate of subjects. On the other hand, there was a statistically significant difference in agreement rates caused by the identity of the AI, with the human-like AI identity leading to lower agreement rates compared to the control group. This finding contradicts the initial expectation that a human-like AI would result in higher agreement.

\section{Discussion}

In this study, we analyzed the impact of cognitive biases on human-AI collaboration by conducting an experiment that replicated the recruitment process of a company. The experiment aimed to investigate the effect of framing bias and anthropomorphism on hiring managers’ decisions when they are assisted by AI recommendations. By gaining a deeper understanding of these biases, and their impact on human-AI collaboration, our goal was to provide valuable insights that can inform the development of AI-based tools, and ultimately enhance the synergy between humans and AI systems in decision-making processes.

\subsection{Framing effect}

Although attribute framing has been extensively demonstrated in the literature \citep{freling_when_2014} through numerous studies \citep{davis_contextual_1986-1,duchon_framing_1989,dunegan_framing_1993,leong_role_2017,levin_associative_1987}, the framing of AI recommendations did not significantly affect subjects’ agreement rate in our experiment. A possible explanation could be that the manipulation chosen for the experiment was insufficient to influence subjects in their decisions. While previous studies were able to induce subjects into framing bias through success and failure rates, the manipulation differed in certain aspects. 

For instance, participants in this study received additional information about candidates' education and work experience alongside the AI recommendations. In contrast, other studies that have demonstrated framing effects provided minimal information to subjects. In Levin’s experiment \citeyear{levin_associative_1987-1}, subjects evaluating basketball players only had access to the percentage of shots missed or passed. Likewise, the study of \citet{duchon_framing_1989} simulated an R\&D financial allocation decision where the main information available to subjects was the rate of success or failure of previous projects. The discrepancy between the experimental design of these studies and our own suggests that providing additional information, such as the candidate’s education and work experience, may have a debiasing effect. This could potentially explain the absence of a significant framing effect in our study. 

The hypothetical debiasing effect of additional information in our experiment is consistent with Kahneman’s dual-system theory, specifically, with the “What You See Is All There Is” principle which rules System 1. This principle suggests that people tend to focus on the information that is presented to them. In the case of framing effects, they can be explained by the way that the negative and positive frames capture individuals’ attention, leading them to overlook the implicit counterparts \citep{kahneman_thinking_2011-1}. Although a rate of success equal to 90\% implies a rate of failure equal to 10\% (and vice-versa), this equivalence is not immediately apparent to individuals. Building on this concept, \citet{kreiner_role_2018-1} successfully moderated framing bias by shifting subjects’ attention to the complementary, non-explicit frame. A similar phenomenon may take place in our experiment. Additional information may have redirected subjects’ attention away from the initially presented frame in the AI recommendation, thereby mitigating the framing effect and debiasing subjects’ judgments. 

Since currently humans rarely rely exclusively on AI decision-aid tools for decision-making, this finding might suggest that the framing effect is not a significant factor that interferes with human-AI collaboration for the moment, provided that the decision-makers have additional information on which to base their judgments. Indeed, some research has found no evidence of the framing effect impacting the reliance of humans on decision-aid systems. For instance, in Brown and Jones' (\citeyear{brown_factors_1998-1}) experiment, subjects had to decide which computer system to recommend to their clients based on the scores of some features. Additionally, they were assisted by a decision-aid algorithm, which computed those scores and recommended a specific computer system to subjects. Subjects were told that they could potentially gain or lose an important client. However, the frame of the scenario did not significantly affect the reliance of subjects on the decision-aid system. A different experiment conducted by \citep{huerta_framing_2012} on assisted decision-making suggests that cultural factors may determine the effects of framing on decision-aid reliance. For the American sample, the framing of a report generated by an automated system did not influence the intention to investigate fraud, whereas it did influence the Mexican sample.

Although AI systems are increasingly present in our society, the framing effect has not yet emerged as a major concern. The lack of studies on this topic serves as proof. However, as AI applications continue to expand, the potential effects of framing bias in AI-assisted decision-making could become increasingly relevant. In this experiment, the AI recommendation was one of the several factors presented to candidates. The additional information provided may have shifted subjects’ attention from the frame, mitigating bias. Nevertheless, it is possible to imagine a future where AI systems are designed to omit external information, or to display solely AI-generated information, potentially introducing framing bias. 

\subsection{Anthropomorphism and AI identities}

The findings of this study reveal a significant effect of the human AI identity on agreement rates among subjects who successfully completed the two attention checks of the survey. Nevertheless, the observed effect was not expected. The human identity was found to have a negative impact on agreement with AI recommendations. Subjects exposed to this identity had a significantly lower agreement rate in comparison to the control group. Consequently, the results of the experiment deviate from the results of the reference paper, where a human image increased reliance on automated decision-aid systems \citep{pak_decision_2012}. 

One potential explanation for the divergence in results may lie in the differences between both manipulations. In contrast to the reference study, which used an image of a person dressed as a doctor, the generic human image used in our experiment did not convey a specific professional affiliation, such as someone working in a company’s HR department. Moreover, the structure of the recommendation also differed between studies. While our study’s recommendation was presented in terms of probabilities, the other study employed what could be considered a more human-like structure (e.g., “my advice is: cholesterol”). This potential explanation would imply that in order to increase agreement with an AI tool, the credibility of AI’s human identity is something that should be considered, especially regarding how it conveys information to the decision-maker.

Another potential explanation for the observed effect of AI identities on subjects’ agreement with the recommendation, and its deviation from the reference paper, is the different nature of the interaction. In the case of a decision-aid system that assists patients in making diabetes management decisions, the system is replacing a human in an already existent and social interaction. It is common to go to the doctor to get assistance on how to better manage our health. On the contrary, in our study, the AI had a task-oriented interaction with subjects in a nonsocial setting. As \citet{yang_anthropomorphism_2022} argue, when the interaction occurs in nonsocial scenarios, AI products are treated as utilities, resulting in an indifference toward its degree of anthropomorphism. Consequently, contrary to the belief that human-like AI identities facilitate human-AI interaction, in certain cases, designers of AI tools should not focus their efforts on anthropomorphism \citep{yang_anthropomorphism_2022}. The findings of our experiment align with this explanation, suggesting that a more neutral identity would be preferable in certain cases.

\subsection{Lack of manipulation checks}

An important limitation of this study, which should be addressed, is the absence of manipulation checks. In the case of AI identities, the study did not measure subjects’ perceived humanness. Consequently, it is unclear whether participants actually perceived the human-like identity as more humanlike compared to the generic or the robot-like conditions. In other words, we cannot determine if subjects anthropomorphized the AI. Therefore, the observed differences cannot be conclusively attributed to perceptions of humanness, but rather to the specific manipulations employed. For instance, differences between the generic condition and the humanlike condition may have been due to the characteristics of the image employed, rather than the degree of humanness. Similarly, the study did not measure the perceived positive or negative sentiment of the framing conditions. As mentioned above, the absence of a significant framing effect might be explained by the failure of the experimental manipulation to induce framing bias in subjects. This lack of measurement is a significant limitation of the study and should be considered. 

\subsection{Conclusions}

The findings of this study have clear practical implications, as AI-based tools are increasingly adopted by companies for various decision-making processes, such as recruitment. Companies are already relying on AI-driven tools to screen candidates’ resumes, optimize their hiring decisions, and even provide assistance to candidates via AI-powered chatbots. In this context, this study has some significant consequences for the development of AI-based tools. 

Although the latest studies have shown that collaborating with AI tools can increase human productivity, the nature of this collaboration remains unclear. Notably, the role of human perception has been largely overlooked so far. This study provides evidence about the fact that cognitive biases can impact human-AI collaboration, ultimately affecting the level of human agreement toward AI recommendations. Consequently, AI tools designed to support human decision-making may produce unintended consequences if the nuances of human-AI collaboration and its effects on human cognition are not fully understood. 

This work also provides evidence about the fact that there is no such thing as a “one size fits all” approach for AI product design. Instead, the findings suggest that multiple aspects should be considered when designing AI tools, such as the situation in which the interaction takes place, the availability of external information, or how the AI output is displayed. Likewise, cultural differences may increase the complexity of worldwide AI product implementations. 

All in all, this work points to the need to further investigate the role of cognitive biases in human-AI collaboration in order to design AI products that effectively interact with humans, while avoiding ethical pitfalls. The results suggest that, on certain occasions, humans are more likely to agree with the AI when its design lacks a specific set of human-like characteristics. In addition, subtle framing manipulations of AI outputs do not seem to significantly affect human decisions. However, the lack of studies in this area and the main limitations of this work, including its inability to replicate framing effects, the large number of subjects who failed the attention checks, and the lack of manipulation checks, indicate that there is still much to be discovered in the realm of human-AI collaboration. By continuing to explore this dynamic, we can develop AI systems that are better designed to work collaboratively with humans in a variety of contexts.  

\section*{Acknowledgements}

We thank Michele Piazzai for constructive feedback on previous versions of this article and generous guidance.

\bibliographystyle{apalike}
\bibliography{ref}

\begin{thebibliography}{}

\bibitem[Arkes, 1991]{arkes_costs_1991}
Arkes, H.~R. (1991).
\newblock Costs and benefits of judgment errors: {Implications} for debiasing.
\newblock {\em Psychological Bulletin}, 110(3):486--498.

\bibitem[Aronow et~al., 2019]{aronow_note_2019}
Aronow, P.~M., Baron, J., and Pinson, L. (2019).
\newblock A {Note} on {Dropping} {Experimental} {Subjects} who {Fail} a {Manipulation} {Check}.
\newblock {\em Political Analysis}, 27(4):572--589.

\bibitem[Bower et~al., 2017]{bower_fair_2017}
Bower, A., Kitchen, S.~N., Niss, L., Strauss, M.~J., Vargas, A., and Venkatasubramanian, S. (2017).
\newblock Fair pipelines.
\newblock arXiv:1707.00391 [cs, stat].

\bibitem[Brown and Jones, 1998]{brown_factors_1998-1}
Brown, D.~L. and Jones, D.~R. (1998).
\newblock Factors that influence reliance on decision aids: a model and an experiment.
\newblock {\em Journal of Information Systems}, 12:75--94.

\bibitem[Buda and Charnov, 2003]{buda_message_2003}
Buda, R. and Charnov, B.~H. (2003).
\newblock Message processing in realistic recruitment practices.
\newblock {\em Journal of Managerial Issues}, 15(3):302--316.

\bibitem[Chinmulgund et~al., 2023]{chinmulgund_anthropomorphism_2023}
Chinmulgund, A., Khatwani, R., Tapas, P., Shah, P., and Sekhar, R. (2023).
\newblock Anthropomorphism of {AI} based chatbots by users during communication.
\newblock In {\em 2023 3rd {International} {Conference} on {Intelligent} {Technologies} ({CONIT})}, pages 1--6, Hubli, India. IEEE.

\bibitem[Dacey, 2017]{dacey_anthropomorphism_2017}
Dacey, M. (2017).
\newblock Anthropomorphism as cognitive bias.
\newblock {\em Philosophy of Science}, 84(5):1152--1164.

\bibitem[Dastin, 2018]{dastin_amazon_2018}
Dastin, J. (2018).
\newblock Amazon scraps secret {AI} recruiting tool that showed bias against women.
\newblock \textit{Reuters}. Retrieved December 26, 2023, from https://www.reuters.com/article/idUSKCN1MK0AG/.

\bibitem[Davis and Bobko, 1986]{davis_contextual_1986-1}
Davis, M.~A. and Bobko, P. (1986).
\newblock Contextual effects on escalation processes in public sector decision making.
\newblock {\em Organizational Behavior and Human Decision Processes}, 37(1):121--138.

\bibitem[De~Visser et~al., 2016]{de_visser_almost_2016}
De~Visser, E.~J., Monfort, S.~S., McKendrick, R., Smith, M. A.~B., McKnight, P.~E., Krueger, F., and Parasuraman, R. (2016).
\newblock Almost human: {Anthropomorphism} increases trust resilience in cognitive agents.
\newblock {\em Journal of Experimental Psychology: Applied}, 22(3):331--349.

\bibitem[Dhimant, 2020]{noauthor_campus_2020}
Dhimant, G. (2020).
\newblock Campus {Recruitment}.
\newblock Retrieved October 26, 2023, from https://www.kaggle.com/datasets/benroshan/factors-affecting-campus-placement.

\bibitem[Duchon et~al., 1989]{duchon_framing_1989}
Duchon, D., Dunegan, K., and Barton, S. (1989).
\newblock Framing the problem and making decisions: The facts are not enough.
\newblock {\em IEEE Transactions on Engineering Management}, 36(1):25--27.

\bibitem[Duffy, 2003]{duffy_anthropomorphism_2003}
Duffy, B.~R. (2003).
\newblock Anthropomorphism and the social robot.
\newblock {\em Robotics and Autonomous Systems}, 42(3-4):177--190.

\bibitem[Dunegan, 1993]{dunegan_framing_1993}
Dunegan, K.~J. (1993).
\newblock Framing, cognitive modes, and image theory: {Toward} an understanding of a glass half full.
\newblock {\em Journal of Applied Psychology}, 78(3):491--503.

\bibitem[Evans and Stanovich, 2013]{evans_dual-process_2013}
Evans, J. S. B.~T. and Stanovich, K.~E. (2013).
\newblock Dual-process theories of higher cognition: {Advancing} the debate.
\newblock {\em Perspectives on Psychological Science}, 8(3):223--241.

\bibitem[Fernández-Martínez and Fernández, 2020]{fernandez-martinez_ai_2020}
Fernández-Martínez, C. and Fernández, A. (2020).
\newblock {AI} and recruiting software: {Ethical} and legal implications.
\newblock {\em Paladyn, Journal of Behavioral Robotics}, 11(1):199--216.

\bibitem[Fink, 2012]{fink_anthropomorphism_2012}
Fink, J. (2012).
\newblock Anthropomorphism and {Human} {Likeness} in the {Design} of {Robots} and {Human}-{Robot} {Interaction}.
\newblock In Ge, S.~S., Khatib, O., Cabibihan, J.-J., Simmons, R., and Williams, M.-A., editors, {\em Social {Robotics}}, pages 199--208, Berlin, Heidelberg. Springer Berlin Heidelberg.

\bibitem[Freling et~al., 2014]{freling_when_2014}
Freling, T.~H., Vincent, L.~H., and Henard, D.~H. (2014).
\newblock When not to accentuate the positive: {Re}-examining valence effects in attribute framing.
\newblock {\em Organizational Behavior and Human Decision Processes}, 124(2):95--109.

\bibitem[Frisch, 1993]{frisch_reasons_1993}
Frisch, D. (1993).
\newblock Reasons for framing effects.
\newblock {\em Organizational Behavior and Human Decision Processes}, 54(3):399--429.

\bibitem[Grimm et~al., 2018]{grimm_team_2018}
Grimm, D.~A., Demir, M., Gorman, J.~C., and Cooke, N.~J. (2018).
\newblock Team situation awareness in human-autonomy teaming: {A} systems level approach.
\newblock {\em Proceedings of the Human Factors and Ergonomics Society Annual Meeting}, 62(1):149--149.

\bibitem[Hjeij and Vilks, 2023]{hjeij_brief_2023}
Hjeij, M. and Vilks, A. (2023).
\newblock A brief history of heuristics: How did research on heuristics evolve?
\newblock {\em Humanities and Social Sciences Communications}, 10(1):64.

\bibitem[Hodgkinson et~al., 2023]{hodgkinson_heuristics_2023}
Hodgkinson, G.~P., Burkhard, B., Foss, N.~J., Grichnik, D., Sarala, R.~M., Tang, Y., and Van~Essen, M. (2023).
\newblock The heuristics and biases of top managers: {Past}, present, and future.
\newblock {\em Journal of Management Studies}, 60(5):1033--1063.

\bibitem[Hoff and Bashir, 2015]{hoff_trust_2015}
Hoff, K.~A. and Bashir, M. (2015).
\newblock Trust in automation: {Integrating} empirical evidence on factors that influence trust.
\newblock {\em Human Factors: The Journal of the Human Factors and Ergonomics Society}, 57(3):407--434.

\bibitem[Huerta et~al., 2012]{huerta_framing_2012}
Huerta, E., Glandon, T., and Petrides, Y. (2012).
\newblock Framing, decision-aid systems, and culture: {Exploring} influences on fraud investigations.
\newblock {\em International Journal of Accounting Information Systems}, 13(4):316--333.

\bibitem[Hunkenschroer and Luetge, 2022]{hunkenschroer_ethics_2022}
Hunkenschroer, A.~L. and Luetge, C. (2022).
\newblock Ethics of {AI}-{Enabled} {Recruiting} and {Selection}: {A} {Review} and {Research} {Agenda}.
\newblock {\em Journal of Business Ethics}, 178(4):977--1007.

\bibitem[Jarrahi, 2018]{jarrahi_artificial_2018}
Jarrahi, M.~H. (2018).
\newblock Artificial intelligence and the future of work: {Human}-{AI} symbiosis in organizational decision making.
\newblock {\em Business Horizons}, 61(4):577--586.

\bibitem[Jin et~al., 2017]{jin_how_2017-1}
Jin, J., Zhang, W., and Chen, M. (2017).
\newblock How consumers are affected by product descriptions in online shopping: {Event}-related potentials evidence of the attribute framing effect.
\newblock {\em Neuroscience Research}, 125:21--28.

\bibitem[Kahneman, 2011]{kahneman_thinking_2011-1}
Kahneman, D. (2011).
\newblock {\em Thinking, fast and slow.}
\newblock Thinking, fast and slow. Farrar, Straus and Giroux.
\newblock Pages: 499.

\bibitem[Kahneman and Frederick, 2002]{gilovich_representativeness_2002}
Kahneman, D. and Frederick, S. (2002).
\newblock Representativeness revisited: {Attribute} substitution in intuitive judgment.
\newblock In Gilovich, T., Griffin, D., and Kahneman, D., editors, {\em Heuristics and biases: {The} psychology of intuitive judgment}, pages 49--81. Cambridge University Press, 1 edition.

\bibitem[Kaur et~al., 2019]{kaur_building_2019}
Kaur, H., Williams, A.~C., and Lasecki, W.~S. (2019).
\newblock Building shared mental models between humans and {AI} for effective collaboration.
\newblock In {\em {CHI}’19, {May} 2019, {Glasgow}, {Scotland}}.

\bibitem[Kim-Schmid and Roshni, 2022]{kim-schmid_where_2022}
Kim-Schmid, J. and Roshni, R. (2022).
\newblock Where {AI} can - and can't - help talent management.
\newblock \textit{Harvard Business Review}. Retrieved November 28, 2023, from https://hbr.org/2022/10/where-ai-can-and-cant-help-talent-management.

\bibitem[Kirk, 2009]{millsap_experimental_2009}
Kirk, R.~E. (2009).
\newblock Experimental design.
\newblock In Millsap, R. and Maydeu-Olivares, A., editors, {\em The {SAGE} {Handbook} of {Quantitative} {Methods} in {Psychology}}, pages 23--46. SAGE Publications Ltd.

\bibitem[Kreiner and Gamliel, 2018]{kreiner_role_2018-1}
Kreiner, H. and Gamliel, E. (2018).
\newblock The {Role} of {Attention} in {Attribute} {Framing}.
\newblock {\em Journal of Behavioral Decision Making}, 31(3):392--401.
\newblock \_eprint: https://onlinelibrary.wiley.com/doi/pdf/10.1002/bdm.2067.

\bibitem[Kreiner and Gamliel, 2019]{kreiner_alive_2019}
Kreiner, H. and Gamliel, E. (2019).
\newblock “{Alive}” or “not dead”: {The} contribution of descriptors to attribute-framing bias.
\newblock {\em Quarterly Journal of Experimental Psychology}, 72(12):2776--2787.

\bibitem[Leong et~al., 2017]{leong_role_2017}
Leong, L.~M., McKenzie, C. R.~M., Sher, S., and Müller‐Trede, J. (2017).
\newblock The role of inference in attribute framing effects.
\newblock {\em Journal of Behavioral Decision Making}, 30(5):1147--1156.

\bibitem[Levin, 1987a]{levin_associative_1987}
Levin, I.~P. (1987a).
\newblock Associative effects of information framing.
\newblock {\em Bulletin of the Psychonomic Society}, 25(2):85--86.

\bibitem[Levin, 1987b]{levin_associative_1987-1}
Levin, I.~P. (1987b).
\newblock Associative effects of information framing on human judgments.
\newblock {\em Paper presented at the annual meeting of the Midwestern Psychological Association}.

\bibitem[Levin et~al., 1998]{levin_all_1998}
Levin, I.~P., Schneider, S.~L., and Gaeth, G.~J. (1998).
\newblock All frames are not created equal: {A} typology and critical analysis of framing effects.
\newblock {\em Organizational Behavior and Human Decision Processes}, 76(2):149--188.

\bibitem[Lukyanenko et~al., 2022]{lukyanenko_trust_2022}
Lukyanenko, R., Maass, W., and Storey, V.~C. (2022).
\newblock Trust in artificial intelligence: {From} a {Foundational} {Trust} {Framework} to emerging research opportunities.
\newblock {\em Electronic Markets}, 32(4):1993--2020.

\bibitem[{Miranda Bogen} and {Aaron Rieke}, 2018]{miranda_bogen_help_2018}
{Miranda Bogen} and {Aaron Rieke} (2018).
\newblock Help wanted: {An} examination of hiring algorithms, equity, and bias.
\newblock Technical report, Upturn.

\bibitem[Mori, 1970]{mori_uncanny_1970}
Mori, M. (1970).
\newblock The uncanny valley.
\newblock {\em Energy}, 7:33--35.

\bibitem[Moussawi et~al., 2021]{moussawi_how_2021}
Moussawi, S., Koufaris, M., and Benbunan-Fich, R. (2021).
\newblock How perceptions of intelligence and anthropomorphism affect adoption of personal intelligent agents.
\newblock {\em Electronic Markets}, 31(2):343--364.

\bibitem[Neth and Gigerenzer, 2015]{scott_heuristics_2015}
Neth, H. and Gigerenzer, G. (2015).
\newblock Heuristics: {Tools} for an uncertain world.
\newblock In Scott, R.~A. and Kosslyn, S.~M., editors, {\em Emerging trends in the social and behavioral sciences: {An} interdisciplinary, searchable, and linkable resource}, pages 1--18. Wiley, 1 edition.

\bibitem[OpenAI, 2022]{openai_introducing_2022}
OpenAI (2022).
\newblock Introducing {ChatGPT}.
\newblock Retrieved October 30, 2023, from https://openai.com/blog/chatgpt.

\bibitem[Pak et~al., 2012]{pak_decision_2012}
Pak, R., Fink, N., Price, M., Bass, B., and Sturre, L. (2012).
\newblock Decision support aids with anthropomorphic characteristics influence trust and performance in younger and older adults.
\newblock {\em Ergonomics}, 55(9):1059--1072.

\bibitem[R.~Chow, 2023]{r_chow_how_2023}
R.~Chow, A. (2023).
\newblock How {ChatGPT} managed to grow faster than {TikTok} or {Instagram}.
\newblock \textit{Time}. Retrieved November 22, 2023, from https://time.com/6253615/chatgpt-fastest-growing/.

\bibitem[Rastogi et~al., 2022]{rastogi_deciding_2022}
Rastogi, C., Zhang, Y., Wei, D., Varshney, K.~R., Dhurandhar, A., and Tomsett, R. (2022).
\newblock Deciding fast and slow: {The} role of cognitive biases in {AI}-assisted decision-making.
\newblock {\em Proceedings of the ACM on Human-Computer Interaction}, 6(CSCW1):1--22.

\bibitem[Schanke et~al., 2021]{schanke_estimating_2021}
Schanke, S., Burtch, G., and Ray, G. (2021).
\newblock Estimating the impact of “humanizing” customer service chatbots.
\newblock {\em Information Systems Research}, 32(3):736--751.

\bibitem[Schmitt and Kim, 2007]{schmitt_selection_2007}
Schmitt, N. and Kim, B. (2007).
\newblock Selection decision-making.
\newblock In Boxall, P., Purcell, J., and Wright, P., editors, {\em The {Oxford} {Handbook} of {Human} {Resource} {Management}}, pages 300--323. Oxford University Press.

\bibitem[Shneiderman, 2020]{shneiderman_bridging_2020}
Shneiderman, B. (2020).
\newblock Bridging the {Gap} {Between} {Ethics} and {Practice}: {Guidelines} for {Reliable}, {Safe}, and {Trustworthy} {Human}-centered {AI} {Systems}.
\newblock {\em ACM Transactions on Interactive Intelligent Systems}, 10(4):1--31.

\bibitem[Sidlauskiene et~al., 2023]{sidlauskiene_correction_2023-1}
Sidlauskiene, J., Joye, Y., and Auruskeviciene, V. (2023).
\newblock Correction: {AI}-based chatbots in conversational commerce and their effects on product and price perceptions.
\newblock {\em Electronic Markets}, 33(1):44.

\bibitem[Silva and Vasconcellos, 2023]{silva_human_2023}
Silva, H. K. D. C.~S. and Vasconcellos, L. (2023).
\newblock Human and algorithmic decision-making in the personnel selection process: {A} comparative bibliometric on bias.
\newblock {\em COLLNET Journal of Scientometrics and Information Management}, 17(1):175--189.

\bibitem[Smith and Cribbie, 2021]{smith_factorial_2021}
Smith, C.~E. and Cribbie, R. (2021).
\newblock Factorial {ANOVA} with unbalanced data: {A} fresh look at the types of sums of squares.
\newblock {\em Journal of Data Science}, 12(3):385--404.

\bibitem[Sowa and Przegalinska, 2020]{sowa_digital_2020}
Sowa, K. and Przegalinska, A. (2020).
\newblock Digital {Coworker}: {Human}-{AI} {Collaboration} in {Work} {Environment}, on the {Example} of {Virtual} {Assistants} for {Management} {Professions}.
\newblock In {\em Digital {Transformation} of {Collaboration}: {Proceedings} of the 9th {International} {COINs} {Conference}}, Springer {Proceedings} in {Complexity}. Springer International Publishing, Cham.

\bibitem[Sowa et~al., 2021]{sowa_cobots_2021}
Sowa, K., Przegalinska, A., and Ciechanowski, L. (2021).
\newblock Cobots in knowledge work: {Human}-{AI} collaboration in managerial professions.
\newblock {\em Journal of Business Research}, 125:135--142.

\bibitem[Troshani et~al., 2021]{troshani_we_2021}
Troshani, I., Rao~Hill, S., Sherman, C., and Arthur, D. (2021).
\newblock Do we trust in {AI}? {Role} of anthropomorphism and intelligence.
\newblock {\em Journal of Computer Information Systems}, 61(5):481--491.

\bibitem[Tversky and Kahneman, 1971]{tversky_belief_1971}
Tversky, A. and Kahneman, D. (1971).
\newblock Belief in the law of small numbers.
\newblock {\em Psychological Bulletin}, 76(2):105--110.

\bibitem[Tversky and Kahneman, 1974]{tversky_judgment_1974}
Tversky, A. and Kahneman, D. (1974).
\newblock Judgment under uncertainty: {Heuristics} and biases.
\newblock {\em Science}, 185:1124--1131.

\bibitem[Tversky and Kahneman, 1981]{tversky_framing_1981}
Tversky, A. and Kahneman, D. (1981).
\newblock The framing of decisions and the psychology of choice.
\newblock {\em Science}, 211(4481):453--458.

\bibitem[Van~Esch and Black, 2019]{van_esch_factors_2019}
Van~Esch, P. and Black, J.~S. (2019).
\newblock Factors that influence new generation candidates to engage with and complete digital, {AI}-enabled recruiting.
\newblock {\em Business Horizons}, 62(6):729--739.

\bibitem[Wang et~al., 2020]{wang_human_human_2020}
Wang, D., Churchill, E., Maes, P., Fan, X., Shneiderman, B., Shi, Y., and Wang, Q. (2020).
\newblock From human-human collaboration to human-{AI} collaboration: {Designing} {AI} systems that can work together with people.
\newblock In {\em Extended {Abstracts} of the 2020 {CHI} {Conference} on {Human} {Factors} in {Computing} {Systems}}, pages 1--6, Honolulu HI USA. ACM.

\bibitem[Waytz et~al., 2014]{waytz_mind_2014-1}
Waytz, A., Heafner, J., and Epley, N. (2014).
\newblock The mind in the machine: {Anthropomorphism} increases trust in an autonomous vehicle.
\newblock {\em Journal of Experimental Social Psychology}, 52:113--117.

\bibitem[Wilson et~al., 2005]{wilson_recognizing_2005-1}
Wilson, T., Wiebe, J., and Hoffmann, P. (2005).
\newblock Recognizing contextual polarity in phrase-level sentiment analysis.
\newblock In Mooney, R., Brew, C., Chien, L.-F., and Kirchhoff, K., editors, {\em Proceedings of {Human} {Language} {Technology} {Conference} and {Conference} on {Empirical} {Methods} in {Natural} {Language} {Processing}}, pages 347--354, Vancouver, British Columbia, Canada. Association for Computational Linguistics.

\bibitem[Xu, 2019]{xu_toward_2019}
Xu, W. (2019).
\newblock Toward human-centered {AI}: a perspective from human-computer interaction.
\newblock {\em Interactions}, 26(4):42--46.

\bibitem[Yang et~al., 2022]{yang_anthropomorphism_2022}
Yang, Y., Liu, Y., Lv, X., Ai, J., and Li, Y. (2022).
\newblock Anthropomorphism and customers’ willingness to use artificial intelligence service agents.
\newblock {\em Journal of Hospitality Marketing \& Management}, 31(1):1--23.

\bibitem[Yarger et~al., 2019]{yarger_algorithmic_2019}
Yarger, L., Cobb~Payton, F., and Neupane, B. (2019).
\newblock Algorithmic equity in the hiring of underrepresented {IT} job candidates.
\newblock {\em Online Information Review}, 44(2):383--395.

\end{thebibliography}

\newpage
\appendix

\section{Appendix}
\label{appendix}
\begin{figure}[h]
    \centering
    \includegraphics[width=0.5\columnwidth]{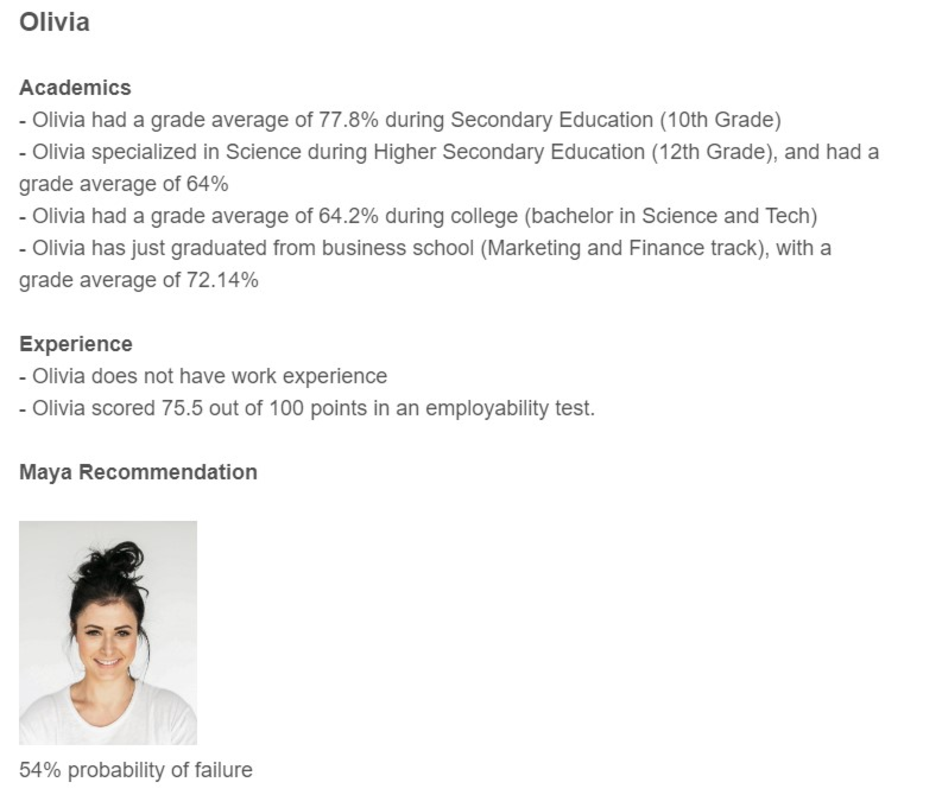}
    \caption{Example of a subject exposed to negative framing (54\% probability of failure), and human-like AI (Maya).}
    \label{human-negative}
    \vspace{0.2in} 
    \includegraphics[width=0.5\columnwidth]{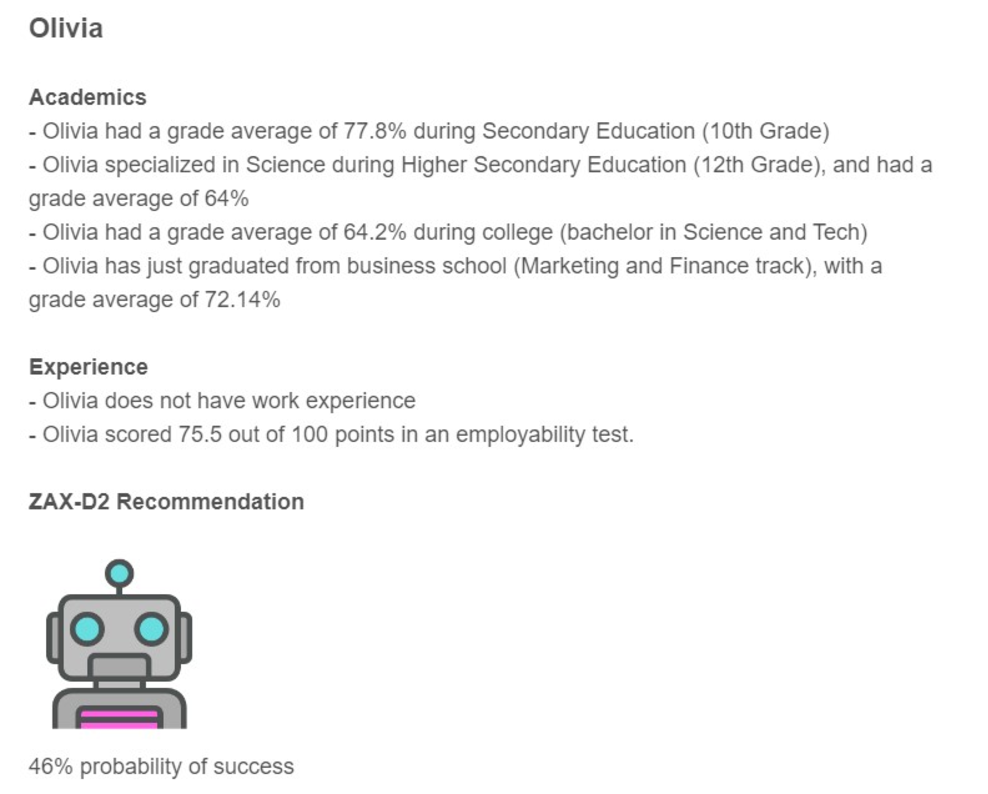}
    \caption{Slide for the same candidate as, but with positive framing (46\% probability of success) and robot-like AI (ZAX-D2).}
    \label{robot-positive}
\end{figure}

\end{document}